Is collaboration among scientists related to the citation impact of papers because their quality increases with collaboration? An analysis based on data from F1000Prime and normalized citation scores

Lutz Bornmann

Division for Science and Innovation Studies

Administrative Headquarters of the Max Planck Society

Hofgartenstr. 8,

80539 Munich, Germany.

Email: bornmann@gv.mpg.de

**Abstract**

In recent years, the relationship of collaboration among scientists and the citation impact of papers have been frequently investigated. Most of the studies show that the two variables are closely related: an increasing collaboration activity (measured in terms of number of authors, number of affiliations, and number of countries) is associated with an increased citation impact. However, it is not clear whether the increased citation impact is based on the higher quality of papers which profit from more than one scientist giving expert input or other (citation-specific) factors. Thus, the current study addresses this question by using two comprehensive datasets with publications (in the biomedical area) including quality assessments by experts (F1000Prime member scores) and citation data for the publications. The study is based on nearly 10,000 papers. Robust regression models are used to investigate the relationship between number of authors, number of affiliations, and number of countries, respectively, and citation impact – controlling for the papers' quality (measured by F1000Prime expert ratings). The results point out that the effect of collaboration activities on impact is largely independent of the papers' quality. The citation advantage is apparently not quality-related; citation specific factors (e.g. self-citations) seem to be important here.

**Key words**

Collaboration; Co-authorship; Bibliometrics; F1000Prime

# 1   Introduction

Collaboration among scientists has become more the norm in modern science than an exception (Ziman, 2000). It is defined by Sonnenwald (2007) as follows: "Scientific collaboration can be defined as interaction taking place within a social context among two or more scientists that facilitates the sharing of meaning and completion of tasks with respect to a mutually shared, superordinate goal" (p. 645). Increases in the expense of equipment push research towards more collective modes of action, because research budgets are limited in most of the countries worldwide. According to Schneider and Sørensen (2015) research systems of smaller countries, such as Switzerland, Denmark, the Netherlands, and Sweden, can be described as efficient, because they achieve a high publication output per capita by means of frequent transnational collaborations. Furthermore, many real-life problems (e.g. climate change) which are intended to be explained and solved by researchers cannot be handled by the "lonely seeker after truth" but only by cooperating scientists from various disciplines and institutions (Bidault & Hildebrand, 2014; Milojević, 2014).

This study deals with a question in the context of scientific collaborations which has been scarcely addressed before. Previous studies have shown that papers written in collaboration (i.e. having more than one co-author with different affiliations and/or from different countries) receive more citations (on average) than papers which are not based on collaboration activities (see the overview of studies in section 2). However, it is not clear whether the citation advantage is especially related to citation specific factors (e.g. self-citations from more than one author of a paper) or the higher quality of the papers (i.e. profiting from the expert knowledge of many researchers). This study thus tests whether the citation advantage of papers written in collaboration is associated with the higher quality of the papers or not. Here two comprehensive databases are used in which not only field-normalized citation scores for every publication but also assessments of the papers' quality are available.

The quality assessments used in this study are based on the F1000Prime post-publication peer review system of papers from the biomedical area (http://f1000.com/prime). In other words, the F1000Prime dataset is used as an alternative source to assess the quality of papers (besides citation scores). Peer assessments of papers have a long tradition in science; they started in the 17th century (Bornmann, 2011) and are formally rooted in the norm "organized skepticism" formulated by Merton (1973) in the ethos of science. According to this norm scientific claims must be exposed to critical scrutiny by peers before being formally accepted. However, the results of many peer review studies have shown that peer assessments are affected by several biases (e.g. national or gender biases) and a low inter-reviewer reliability (Bornmann, 2011; Weller, 2002). Reviewers have different epistemic views, norms, ideologies, and agendas leading to a low inter-reviewer reliability. Although these and other weaknesses of peer assessments are known since many years, they are still assumed as the best possible method to assess the quality of papers in science (Bence & Oppenheim, 2004). According to Martin and Irvine (1983), quality assessments undertaken by peers can be differentiated from quality assessments on the basis of citation counts by the fact that citations are able to measure one part of quality, namely impact. The other parts are importance and accuracy of research.

## 2      Literature overview and purpose of the study

In scientometrics, numerous papers have been published on collaborations in science. Most of these (empirical) studies deal with collaborations measured by co-authorships. For example, Bornmann, Stefaner, de Moya Anegón, and Mutz (2015) published a web application with visualizations of collaborations in science based on co-authorship data (see www.excellence-networks.net). Other forms of measurable collaborations are jointly submitted grant proposals and co-patent applications (Cimenler, Reeves, & Skvoretz, 2014) which are seldom studied in scientometrics. The most important advantage of measuring collaboration by co-authorships is that one focuses on a definite output (i.e. publications) whose quality can be measured (e.g. in terms of citation counts as one part of quality, see above). Further advantages are "verifiability, stability over time, data availability and ease of measurement" (Bozeman, Fay, & Slade, 2013, p. 2). However, when using co-authorships as proxies of collaboration one should have in mind that co-authorships might either reflect only one part of or overestimate collaboration activities between scientists: it will frequently be the case that a lot of activities involving colleagues (e.g. discussions and joint analyses) are running within a research project, which do not always result in co-authorships. The phenomenon that scientists make substantial contributions, but are not mentioned as co-authors on the paper, is described as ghost-authorship. On the other side, authorship can be in the form of honorary authorship in which a scientist is mentioned who does not play any significant role in the work (e.g. a senior academic puts his/her name on a junior colleague's paper) (Bidault & Hildebrand, 2014).

In the Snowball Metrics Recipe Book, Colledge (2014) defines how collaboration can be measured by co-authorships. The author differentiates between national and international co-authorships and publications which have both national and international co-authorships. Many studies in the field of scientometrics have dealt with the extent of these forms of collaboration in science. Overviews can be found in Katz and Martin (1997), Bozeman et al. (2013) as well as Bidault and Hildebrand (2014). The numbers reported in the empirical studies let Bozeman et al. (2013) come to the conclusion that "there is abundant evidence that research collaboration has become the norm in every field of scientific and technical research" (p. 1). This conclusion is supported by two recent studies investigating collaboration on the base of the entire Web of Science (WoS) database (Thomson Reuters): Larivière, Gingras, Sugimoto, and Tsou (2015) analyzed publication data from 1900 to 2011. They included 28,160,453 papers (articles, notes, and reviews) in natural and medical sciences (NMS) as well as 4,347,229 papers in the social sciences and humanities (SSH). Their main results are as follows:

"From 1900 onward, co-authorship, interinstitutional collaboration, and international collaboration have been increasing in both NMS and SSH. More specifically, single-authored papers decreased in NMS from 87% in 1900 to 7% in 2011 and, in SSH, from 97% to 38% during the same period. For the beginning of the period in question, the decrease in single-authored papers in NMS is due to the increase in papers with two authors; the proportion of the latter has also decreased since the beginning of the 1960s, mainly in favor of papers with more than two authors. Papers with one address have also been decreasing, accounting in 2011 for 32% and 46% of all papers in NMS and SSH, respectively. Hence, for both domains, the majority of contemporary papers are the result of interinstitutional collaboration. However, despite its increase throughout the period, international collaboration remains, at the global level, a relatively marginal phenomenon, in 2011, 22.7% and 16.4% of all papers in NMS and SSH, respectively. Similarly, multilateral collaboration, that is, collaborations involving more than two countries, accounted, in 2011, for only 5.1% of all papers in NMS and 2.8% of all papers in SSH" (Larivière et al., 2015, pp. 1330-1331).

The study of Waltman, Tijssen, and van Eck (2011) focusses on geographical distances between authors of the same publication. Their study is based on a similar comprehensive WoS data set as the study of Larivière et al. (2015). Waltman, Tijssen, et al. (2011) state that "across all countries and fields of science, reveals that contemporary science has globalised at a fairly steady rate during recent decades. The average collaboration distance per publication has increased from 334 km in 1980 to 1553 km in 2009. Despite significant differences in globalisation rates across countries and fields of science, we observe a pervasive process in motion, moving towards a truly interconnected global science system" (p. 574).

Whereas one part of the studies dealing with collaborations investigates their extent, another part focuses on the effects of collaborations: are papers written in collaboration more successful (in terms of citations or acceptance rates at journals) than papers which are not based on collaboration activities? This question and similar questions have been treated on different levels of collaboration (between authors, institutions, and countries). For example, Smith, Weinberger, Bruna, and Allesina (2014) divided "publication success into two categories: journal placement and citation performance. Analyzing all papers published between 1996 and 2012 in eight disciplines, we find that those with more countries in their affiliations performed better in both categories". A positive relationship between the citation impact of papers and the number of authors, affiliations, or countries appearing with a paper has also been reported by Larivière et al. (2015) (see the detailed description of the study above) and by Elsevier and Science Europe (2013) investigating institutional collaborations in Europe and between US states. Adams (2013) summarized the results of studies on the relationship between citation impact and collaboration (measured by bibliometric indicators): "Citation impact is typically greater when research groups collaborate, and the benefit strengthens when co-authorship is international" (p. 559). A similar conclusion has been formulated by Sonnenwald (2007): "Numerous bibliometric studies have illustrated that co-authored papers in all disciplines investigated tend to be published in higher-impact journals, cited more frequently, and cited for longer periods of time" (p. 668).

A third group of studies in the area of scientific collaborations investigated collaborations by using social network techniques. A review of corresponding studies can be found in Kumar (2015). For example, Bornmann, Wagner, and Leydesdorff (2015) generated co-authorship networks among authors of highly cited papers for 1995, 2000, 2005, and 2010 to view changes in the participation of BRICS countries (Brazil, Russia, India, China, and South Africa) in global science. Typically, these studies focus on the investigation of collaborations between certain groups of authors (e.g. in a specific discipline) or certain countries.

This study investigates an aspect of collaboration which has been scarcely targeted up to now. Although many studies have shown that the citation impact of papers increases with increasing collaborations (e.g. more co-authors), it is not clear whether this relationship is rather associated with the higher quality of publications or by other factors (e.g. their better visibility). Sonnenwald (2007) describes both perspectives as follows: "Co-authors contribute different types of knowledge and collaborative work may foster more rigorous review of papers, thus increasing the quality of the final publication. Moreover, coauthors can increase the visibility of a paper when they share information about it in conference and workshop presentations, discuss it informally with colleagues, and distribute preprints to colleagues" (p. 668). Thus, this study investigates whether the quality of papers (measured by assessments of experts in the F1000Prime peer review system) has an effect on the relationship between collaboration activities and citation impact. Does the citation advantage of papers written in collaboration level off if the quality of the papers is controlled for in a regression model? In order to find an answer on this question, two comprehensive datasets are used with normalized citation scores (in-house database based on the WoS) and assessments of experts (F1000Prime).

The author of this study has already used the extensive F1000Prime dataset for the investigation of different research topics: Bornmann (2015a) investigates the reliability and predictive validity of the experts' ratings in the F1000Prime peer review system. Bornmann (2014c, 2015b) used the data to investigate whether alternative metrics (altmetrics, see Bornmann, 2014a) are able to measure the societal impact of papers. Two further papers were also published in the area of altmetrics: Bornmann and Haunschild (2015) and Haunschild and Bornmann (2015) present altmetric statistics for papers in the F1000Prime system.

# 3 Methods

## 3.1 Datasets used

F1000Prime is a post-publication peer review system of papers from medical and biological journals (see http://f1000.com/prime). The system is an information service for the biomedical community which has been available from the Science Navigation Group since 2002. Papers for F1000Prime are selected by a peer-nominated global "Faculty" of leading scientists and clinicians who rate the papers and explain their importance. Since the so called Faculty members can select any paper of interest (i.e. the papers are not systematically selected and rated), only a restricted set of papers from the medical and biological journals is included in F1000Prime (Kreiman & Maunsell, 2011; Wouters & Costas, 2012). However, "the great majority [of Faculty members] pick papers published within the past month, including advance online papers, meaning that users can be made aware of important papers rapidly" (Wets, Weedon, & Velterop, 2003, p. 254). The papers included in F1000Prime are rated by the Faculty members as "Good", "Very good", or "Exceptional" which is equivalent to scores of 1, 2, or 3, respectively. In many cases a paper is not evaluated by one Faculty member alone but by several.

In January 2014, F1000Prime provided the author of this study with data on all ratings made and the bibliographic information for the corresponding papers in their system (n=149,227 records). The dataset contains a total of 104,633 different DOIs which, with a few exceptions, are all individual papers (not all DOIs refer to a specific paper). This sharp reduction from records to DOIs is due to the fact that the F1000Prime dataset was generated on the level of ratings and not publications (many publications have received more than one rating). Since the dataset does not contain any citation impact scores or other bibliometric data, it was matched with a bibliometric in-house database at the Max Planck Society (MPG), which is administered by the Max Planck Digital Library (MPDL) and is based on the WoS.

In order to establish a link between the F1000Prime papers and the bibliometric data, two procedures are followed in this study: a total of 90,436 F1000Prime papers could be matched with a paper in the in-house database using the DOI. (2) For 4,205 of the 14,197 remaining papers, no match was possible with the DOI, but one could be achieved with the name of the first author, the journal, the volume and the issue. Bibliometric data were then available for 94,641 papers of the 104,633 in total (91%). A similar procedure, which matched data from F1000Prime with bibliometric data in another in-house database (based on WoS), led to a similar percentage of 93% (Waltman & Costas, 2014).

For this study, the F1000Prime dataset is reduced to only those papers with at least two ratings of Faculty members. In order to increase the reliability of the quality assessments, papers with only one rating are excluded. Assessments by experts might be personally biased (see above) and the consideration of more than one rating (from different experts) should increase the reliability of the assessments (Bornmann & Marx, 2014). In order to have only one assessment score from the Faculty members for every paper in the dataset, the median is calculated over the members' ratings for one and the same paper.

From the bibliometric in-house database, the following data was downloaded for this study in order to measure the citation impact of papers, the mean normalized citation score (MNCS) is used (Waltman, van Eck, van Leeuwen, Visser, & van Raan, 2011a; Waltman, van Eck, van Leeuwen, Visser, & van Raan, 2011b). Since citation counts are dependent on the subject category to which a given paper was assigned (Marx & Bornmann, 2015) and the year of its publication, the citation counts are normalized by an average citation rate calculated on the base of a suitable reference set: the reference set consists of all papers which were published in the same subject category and publication year as the paper in question. The resulting normalized citation scores are larger or smaller than one, whereby the score of one identifies papers which received a similar citation impact (on average) to that of the corresponding papers in the reference set. The MNCS is a standard indicator in bibliometrics.

To enable a citation window of at least three years for every publication (Glänzel, 2008), only publications with publication years smaller than 2013 are considered in this study.

In order to measure the extent of collaboration, three measures are used on the level of single publications: (1) The number of authors measures the extent of collaborations in total. One can expect that all scientists who substantially contributed to a paper are mentioned as co-authors (see above). However, collaborations measured by co-authorships also reflect collaborations between scientists of the same institution (i.e. without consideration of any geographical parameters). (2) Thus, the number of affiliations used as a measure of collaboration restricts the collaborative activities to those of more than one organizational units (nationally or internationally). (3) International collaboration is measured by the number of countries given on a paper: the more countries are involved in the publication of a paper, the more internationally the research was arranged.

The number of affiliations and the number of countries in the dataset of this paper do not consider multiple mentions: for example, if a paper has five authors with three from Germany and two from Switzerland, the number of countries is equal to two.

The consideration of only those papers published before 2013 with (1) at least two ratings of Faculty members and (2) bibliometric data available in the in-house database reduces the F1000Prime dataset (n=94,641 papers, see above) to n=16,557 papers for this study. These papers were published between 1996 and 2012.

### 3.2 Statistics used

The program Stata is used for the statistical analyses in this study (StataCorp., 2015).

Since the variables used in this study do not follow the normal distribution (tested with the skewness/ kurtosis tests for normality), spearman rank-order correlation coefficients are calculated (Sheskin, 2007). The coefficients are interpreted against the backdrop of guidelines published by Cohen (1988) and Kraemer et al. (2003).

In order to inspect the relationship between collaboration activities and ratings of the Faculty members or normalized citation scores, respectively, several (multiple) linear regression models are performed. These models show how the dependent variable (e.g. the normalized citation score) is related to one (e.g. the number of authors) or two independent variables (e.g. the number of authors and the members' ratings). Thus, the linear regression model stipulates that the dependent variable can be written as

$$y_i = \beta_0 + \beta_1 x_{i1} + \beta_2 x_{i2} + \varepsilon_i, \quad i = 1, 2, 3, \ldots, n \tag{1}$$

where $x_1$ and $x_2$ are two independent variables and $\beta_0$, $\beta_1$, and $\beta_2$ are the regression parameters. Error $\varepsilon_i$ is assumed to follow a normal distribution with mean 0 and constant variance $\sigma^2$.

Regression models show how much variance of the dependent variables can be explained by the independent variables. Further, it can be investigated how an increase in collaboration activities is related to citation impact scores if the quality of the papers is controlled for (i.e. the members' ratings are fixed at their mean value). An important assumption of regression models is that the residuals – the deviations of the dependent variable values from the fitted model function – are normally distributed (see equation 1). Also, the skewness/ kurtosis tests for normality are used to test this for every model. Since the results of all tests show that there are concerns about the distribution of the residuals, robust regressions are run. A robust regression uses a sandwich estimator to estimate the standard errors. That means the variance-covariance matrix of the standard errors is estimated in a way that does not assume normality. This yields different $t$-values for testing the significance of parameter estimates (e.g. the explained variance, $R^2$) (Acock, 2014). In order to inspect the effect of the use of the sandwich estimator on the results, both the results of the classical linear regression models and the robust regression models are presented in the following.

The consideration of independent variables in the regression models (such as the number of authors or affiliations) initially involves some questionable assumptions (Bornmann & Williams, 2013). For example, one assumes that the more authors a paper has, the higher its citation impact. It is probably more reasonable to assume that, after a certain point, additional authors produce less and less additional citation impact. To address such diminishing returns, squared terms for independent variables are added to the regression models. Squared terms allow for the possibility that the independent variables may have a negative effect on the dependent variable (Acock, 2014; Berry & Feldman, 1985). The squared terms are considered in the regression models if the inclusion is theoretically plausible and the resulting coefficients are negative and statistically significant.

Subsequent to the regression model estimations, predicted values are calculated using the margins command in Stata (Bornmann & Williams, 2013; Williams, 2012; Williams & Bornmann, 2014). Predicted values are useful in investigating the substantive and practical significance of findings besides the sign and statistical significance of model parameters. Predicted values are visualized in order to inspect the relationship between dependent and independent variables. Further, the relationship between the dependent and independent variables is visualized while controlling for another independent variable.

# 4    Results

## 4.1    Correlation analyses

In a first step of analysis, all variables considered in this study are correlated in order to provide a first impression of the relationship between the variables. Table 1 shows the Spearman rank-order correlation coefficients for the correlation between number of authors, number of affiliations, number of countries, Faculty members' ratings, and normalized citation scores. If the coefficients in Table 1 are interpreted against the backdrop of the guidelines of Cohen (1988) and Kraemer et al. (2003), they reveal small or smaller than typical correlation coefficients between the ratings of the Faculty members and the numbers of authors, affiliations, and countries, respectively. In other words, this proxy of quality (assessments of experts) does not seem to be related to the number of authors, affiliations, and countries. These features of publications (reflecting collaboration activities) are obviously less relevant for the ratings of the Faculty members.

Table 1. Spearman rank-order correlation coefficients for the correlation between number of authors, number of affiliations, number of countries, Faculty members' ratings, and normalized citation scores ($n$=16,554)*

|  | Authors | Affiliations | Countries | Ratings | Scores |
|---|---|---|---|---|---|
| Authors | 1.00 |  |  |  |  |
| Affiliations | 0.71 | 1.00 |  |  |  |
| Countries | 0.44 | 0.52 | 1.00 |  |  |
| Ratings | 0.09 | 0.06 | 0.07 | 1.00 |  |
| Scores | 0.34 | 0.27 | 0.17 | 0.24 | 1.00 |

Note. * Since citation scores are not available for three publications, the analyses are based on a reduced dataset.

A similar weak relationship is reported by Wouters et al. (2015) who correlated the number of authors with outcomes from the Research Excellence Framework (REF) peer review process. For the grant peer review process the results of Jayasinghe (2003) suggest that "the number of researchers in the team did not influence the proposal ratings (committee and final assessor ratings), individual assessor ratings (project and researcher ratings) or success of the proposal" (p. 174).

Also, Table 1 shows the correlations for the normalized citation scores: in contrast to the ratings of the Faculty members, these scores correlate on a medium or typical level not only with the Faculty members' ratings, but also with the numbers of authors and affiliations. The correlation with the number of countries is on a lower level.

## 4.2 Regression models with Faculty members' ratings as dependent variable

In a second step of analysis, the relationship between the variables is investigated in more detail on the base of regression models. Before we come to the specific issue of this study in the next section (the relationship between citation impact and collaboration activity – with quality being controlled for), the relationship between expert ratings and collaboration activities is studied in this section. The idea behind this analysis is that Faculty members' ratings can only be used as a proxy for quality in this study if they are themselves not related to collaboration activities. If they were related to quality, the expected citation impact advantage of papers written in collaboration would not be a specific impact related phenomenon, but a general feature of quality indicators. In other words, quality could be generally dependent on collaboration activities (because it profits from the impact of many). Further, the relationship between collaboration and citation impact can only be validly measured with quality being controlled for (in the next section), if the indicator used for measuring the quality is itself not related to collaboration activities. Otherwise the interpretation of the results becomes difficult.

With regression models the proportion of variance in Faculty members' ratings can be explained by the number of authors, number of affiliations, and number of countries. Further, the analyses can reveal which quality scores (median ratings of the Faculty members) are expected at different levels of collaboration activities. Table 2 shows the key figures of the dependent and independent variables included in the models. Three models are calculated with median ratings as the dependent variable and number of authors, number of affiliations, or number of countries as independent variable. As expected, the median values in Table 2 show, that the papers have on average more authors than affiliations and more affiliations than countries.

Table 2. Key figures of the dependent and independent variables ($n$=16,557)

|  | Arithmetic average | Median | Standard deviation | Minimum | Maximum |
|---|---|---|---|---|---|
| Dependent variable |  |  |  |  |  |
| Median rating of the Faculty members | 1.64 | 1 | 0.49 | 1 | 3 |
| Independent variables |  |  |  |  |  |
| Number of authors | 8.72 | 7 | 14.39 | 1 | 964 |
| Number of affiliations | 4.39 | 3 | 5.14 | 0 | 220 |
| Number of countries | 1.64 | 1 | 1.29 | 0 | 27 |

The results of the regression models are shown in Table 3. The results reveal that the number of authors, number of affiliations, and number of countries can only explain less than 1% of the variance in the Faculty members' ratings. The coefficients in the table point out that effects of the collaboration activities on the ratings exist, but these effects are very small. According to the guidelines of Acock (2014) the standardized beta coefficients in Table 3 point to a weak correlation between the dependent and independent variables.

Table 3. Coefficients, standardized beta coefficients and $t$ statistics for three regression models with the median ratings of the Faculty members as dependent variable ($n$=16,557). Results of the classical and robust regression models are reported.

|  | Classical regression | | | Robust regression | | |
|---|---|---|---|---|---|---|
|  | Coefficient | beta | $t$ statistics | Coefficient | beta | $t$ statistics |
| (1) Number of authors ($R^2$=0.6%/0.6%)$^\$$ |  |  |  |  |  |  |
| Number of authors | 0.01* | 0.13 | 9.50 | 0.01* | 0.13 | 7.97 |
| Number of authors (squared) | -0.00* | -0.09 | -6.61 | -0.00* | -0.09 | -6.28 |
| (2) Number of affiliations ($R^2$=0.3%/0.3%)$^\$$ |  |  |  |  |  |  |
| Number of affiliations | 0.01 | 0.08 | 6.88 | 0.01 | 0.08 | 6.54 |
| Number of affiliations | -0.00 | -0.04 | -3.38 | -0.00 | -0.04 | -2.79 |
| (squared) |  |  |  |  |  |  |
| (3) Number of countries ($R^2$=0.3%/0.3%)$^\$$ |  |  |  |  |  |  |
| Number of countries | 0.04 | 0.10 | 6.85 | 004 | 0.10 | 6.98 |
| Number of countries (squared) | -0.00 | -0.06 | -4.19 | -0.00 | -0.06 | -4.52 |

Notes. * $p<0.01$
$^\$$ The $R^2$ for the classical regression model is reported on the left side and the $R^2$ for the robust regression model on the right side.

Subsequent to the model estimations, adjusted predictions of the Faculty members' ratings are calculated to visualize the small effects of the collaboration activities. Although an increase of the number of authors, number of affiliations, and number of countries is associated with an increase in the adjusted predictions of the members' ratings in Figure 1, the increase in the adjusted predictions of the ratings is small. Again, an effect of the variables reflecting collaborations is observable, but the effects are very small.

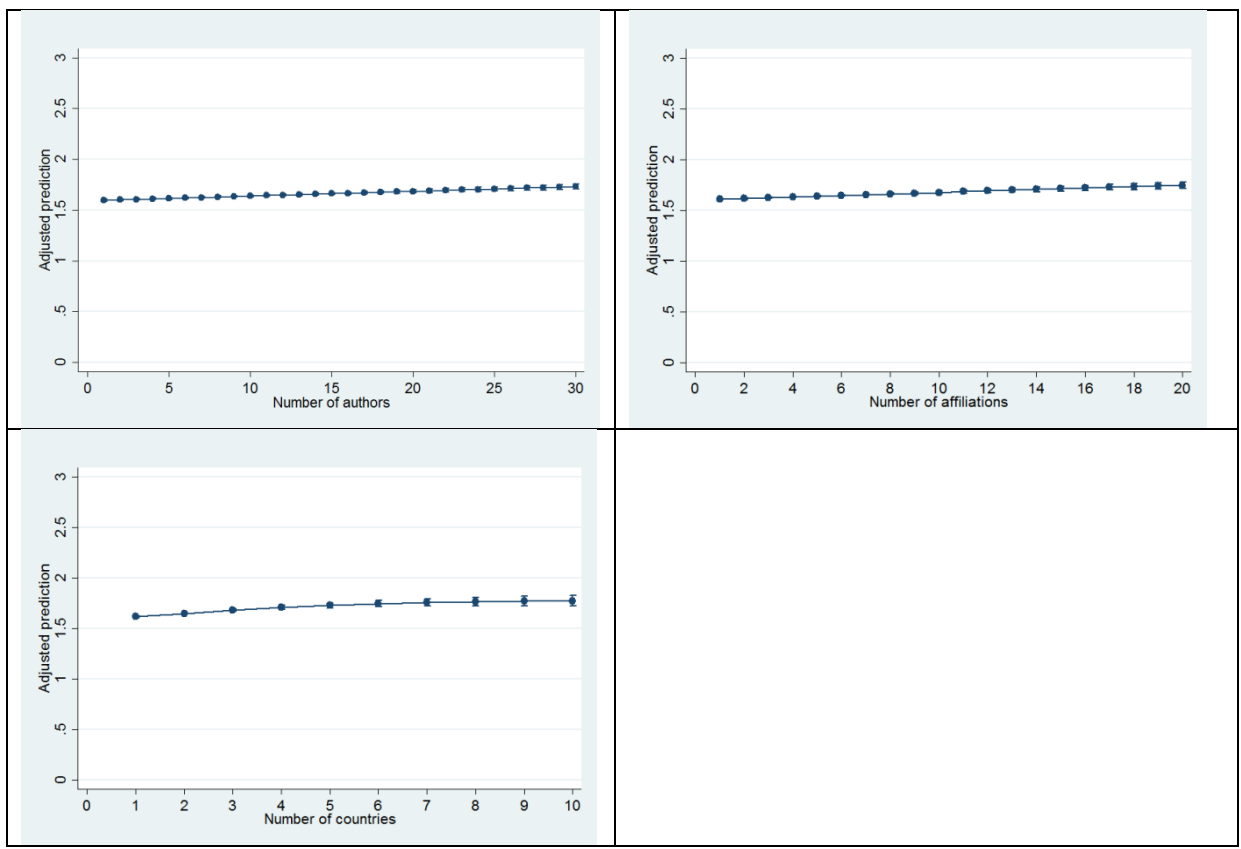

Figure 1. Relationship between number of authors, number of affiliations, and number of countries and adjusted predictions of Faculty members' ratings (based on the results of the robust regressions)

## 4.3 Regression models with normalised citation scores as dependent variable

In a third step of the analysis, the normalized citation scores are considered as dependent variable in the regression models with the Faculty members' ratings as well as the number of authors, number of affiliations, and number of countries included as independent variables. These models are intended to show whether the relationship between collaboration activities and citation scores changes if the quality of the papers is controlled for (in terms of the members' ratings). Table 4 shows the key figures of the dependent and independent variables which are included in six regression models. The median normalized citation score equals 3.44, which reveals a very high citation impact of the papers in the dataset compared to similar papers in the same subject category and publication year.

Table 4. Key figures of the dependent and independent variables (*n*=16,554)*

|  | Arithmetic average | Median | Standard deviation | Minimum | Maximum |
|---|---|---|---|---|---|
| Dependent variable |  |  |  |  |  |
| Normalised citation scores | 5.58 | 3.44 | 8.25 | 0 | 241.49 |
| Independent variables |  |  |  |  |  |
| Median rating of the Faculty members | 1.64 | 1.5 | 0.49 | 1 | 3 |
| Number of authors | 8.72 | 7 | 14.39 | 1 | 964 |
| Number of affiliations | 4.39 | 3 | 5.14 | 0 | 220 |
| Number of countries | 1.64 | 1 | 1.29 | 0 | 27 |

Note. * Since citation scores were not available for three publications, the analyses are based on a reduced dataset.

The results of the regression analyses are shown in Table 5. For the number of authors, the number of affiliations, and the number of countries, two models are estimated: one model considers the median ratings of the Faculty members and the other does not. The table reports standardized beta coefficients where all variables have a mean of 0 and a standard deviation of 1. As the coefficients show, the number of authors and number of affiliations have a moderate effect on citation scores – independent of the inclusion or exclusion of the median Faculty members' ratings. Thus, the effect of quality (measured by the members' ratings) is small. This changes in the model with the number of countries, where the number of countries and the median ratings have a similar weak effect on citation scores.

Table 5. Coefficients, standardized beta coefficients and *t* statistics for six regression models with normalized citation scores as dependent variable (*n*=16,554). Since citation scores were not available for three publications, the analyses are based on a slightly reduced dataset. Each model with the number of authors, the number of affiliations, and the number of countries as independent variables has been calculated twice: excluding (1a, 2a, and 3a) and including (1b, 2b, and 3b) median ratings of the Faculty members. Results of the classical and robust regression models are reported.

|  | Classical regression | | | Robust regression | | |
|---|---|---|---|---|---|---|
|  | Coefficient | beta | *t* statistics | Coefficient | beta | *t* statistics |
| (1a) **Number of authors** ($R^2$=7.7%/7.7%)$^\$$ | | | | | | |
| Number of authors | 0.27* | 0.47 | 35.44 | 0.27* | 0.47 | 10.71 |
| Number of authors (squared) | -0.00* | -0.30 | -22.62 | -0.00* | -0.30 | -9.69 |
| (1b) **Number of authors** ($R^2$=9.8%/9.8%)$^\$$ | | | | | | |
| Number of authors | 0.26* | 0.45 | 34.31 | 0.26* | 0.45 | 10.50 |
| Number of authors (squared) | -0.00* | -0.29 | -21.84 | -0.00* | -0.29 | -9.54 |
| Median ratings | 2.22* | 0.14 | 19.43 | 2.22* | 0.14 | 14.08 |
| (2a) **Number of affiliations** ($R^2$=7.4%/7.4%)$^\$$ | | | | | | |
| Number of affiliations | 0.61* | 0.38 | 34.81 | 0.61* | 0.38 | 13.97 |
| Number of affiliations (squared) | -0.00* | -0.20 | -18.04 | -0.00* | -0.20 | -10.23 |
| (2b) **Number of affiliations** ($R^2$=9.6%/9.6%)$^\$$ | | | | | | |
| Number of | 0.59* | 0.37 | 34.11 | 0.59* | 0.37 | 13.85 |
| affiliations | | | | | | |
| Number of affiliations (squared) | -0.00* | -0.19 | -17.72 | -0.00* | -0.19 | -10.57 |
| Median ratings | 2.30* | 0.15 | 20.14 | 2.30* | 0.15 | 14.85 |
| (2a) **Number of countries** ($R^2$=4.3%/4.2%)$^\$$ | | | | | | |
| Number of countries | 0.94* | 0.15 | 10.56 | 1.31* | 0.21 | 9.89 |
| Number of countries (squared) | 0.04* | 0.07 | 4.96 | § | | |
| (2b) **Number of countries** ($R^2$=6.8%/6.6%)$^\$$ | | | | | | |
| Number of countries | 0.85* | 0.13 | 9.58 | 1.27* | 0.20 | 9.63 |
| Number of countries (squared) | 0.04* | 0.08 | 5.69 | § | | |
| Median ratings | 2.41* | 0.16 | 10.74 | 2.39* | 0.16 | 14.96 |

Note. * $p<0.01$.
§ Since the squared number of countries is not statistically significant, it is excluded from the model.
$ The $R^2$ for the classical regression model is reported on the left side and the $R^2$ for the robust regression model on the right side.

Table 5 reports for each model the explained variance ($R^2$). For example, the results of the regression model including the number of authors show that this independent variable can explain 7.7% of the citation scores' variance. Including the independent variable "median ratings" lead to a small increase of the explained variance: the two variables together explain 9.8%. Similar small increases of the explained variances are visible for the number of affiliations and countries if the ratings are considered. Compared to the results on the relationships between number of authors, number of affiliations, and number of countries, respectively, and adjusted predictions of Faculty members' ratings which showed $R^2$ of less than 1%, here the number of authors, the number of affiliations, and the number of countries can explain a significantly higher proportion of variance (between 4.2% and 7.7%). That means there is a specific effect of these variables on citation scores, which is scarcely visible in another proxy of quality, namely F1000Prime expert ratings.

Figure 2 shows the adjusted predictions resulting from the regression models including and excluding Faculty members' ratings. As the results in the figure show, the adjusted predictions are similarly distributed independent of whether the ratings are included in the models or not (besides the number of authors, the number of affiliations, and the number of countries). The results can be interpreted as follows: in the models including members' ratings, the quality of the papers is controlled for (the median ratings are fixed at their mean value of 1.63). If the (linear) relationship between collaboration activities and citation scores were quality based, this relationship should change (diminish) with the papers' quality being controlled for. Since such a change is scarcely visible in Figure 2, the increase in citation impact with increasing collaboration activities seems to be related to citation-specific factors (such as self-citations) and seems not to be based on quality.

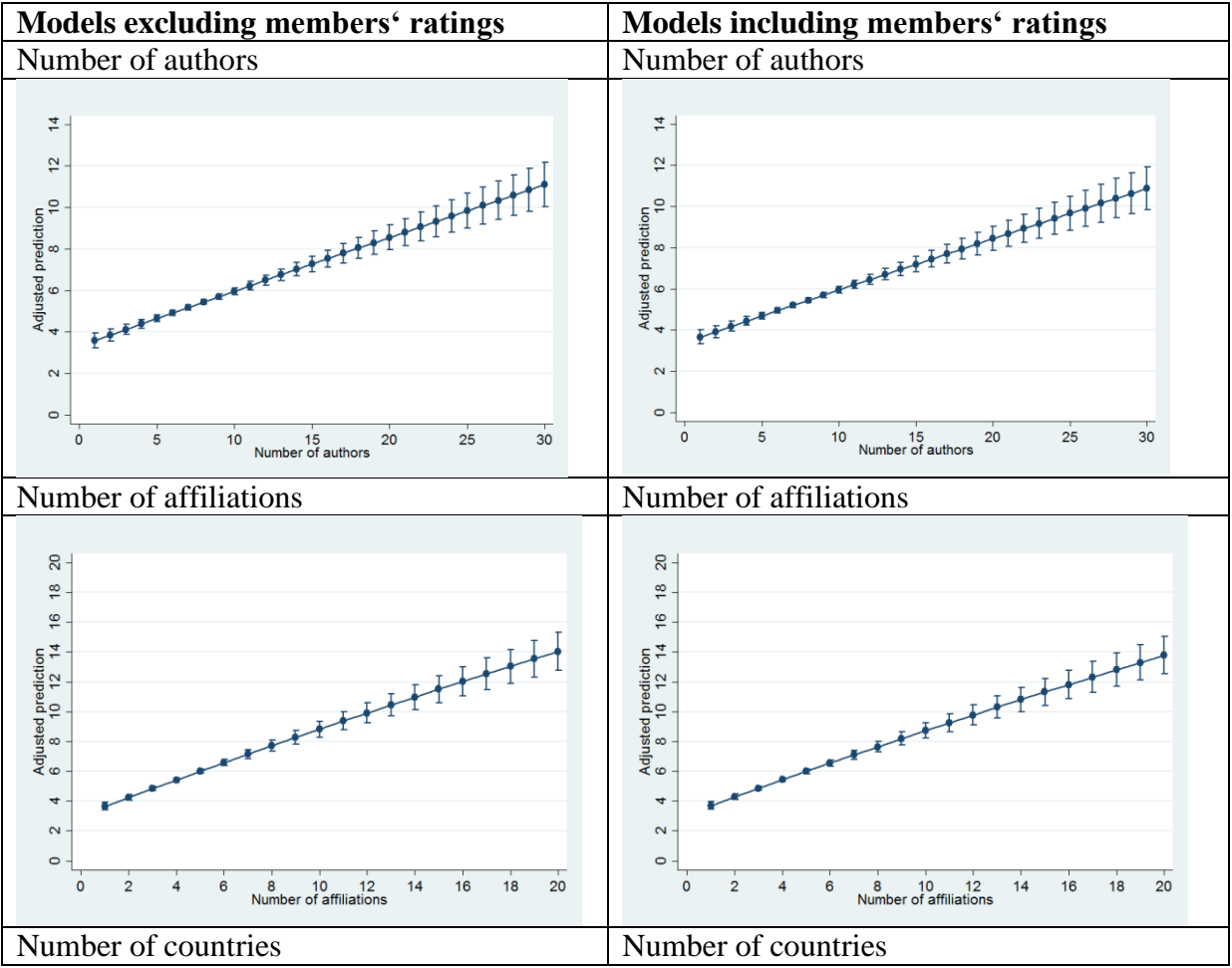

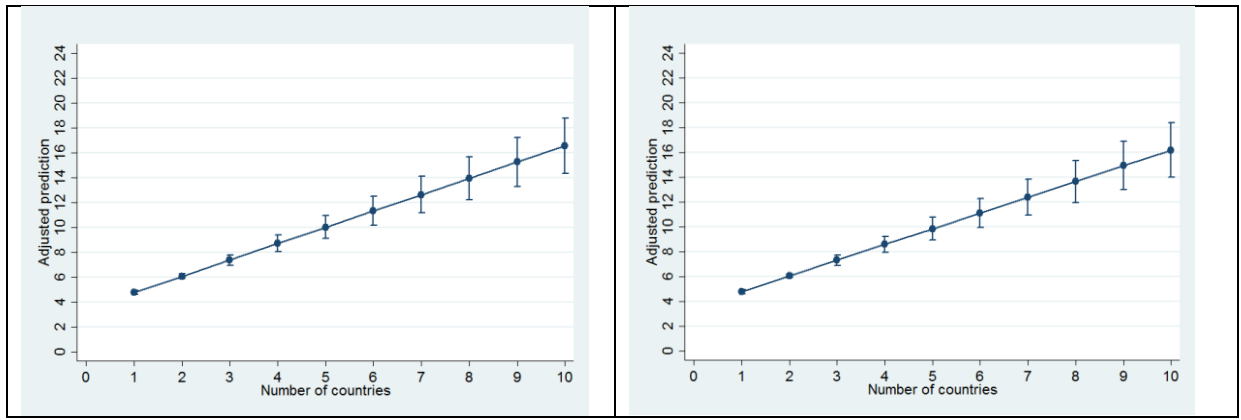

Figure 2. Relationship between number of authors, number of affiliations, and number of countries and normalized citation scores (with and without control for Faculty members' ratings in the regression, based on the results of the robust regressions).

# 5 Discussion

Most of the studies investigating collaboration in science come to the conclusion that this is an increasing phenomenon: "Global collaboration continues to grow as a share of all scientific cooperation, measured as co-authorships of peer-reviewed, published papers. The percent of all scientific papers that are internationally coauthored has more than doubled in 20 years, and they account for all the growth in output among the scientifically advanced countries" (Wagner, Park, & Leydesdorff, 2015). In general, scientists profit from collaboration in research projects. According to Adams (2012) "knowledge is better transferred and combined by collaboration" (p. 335).

In a first step of the analysis in this study, the relationship between collaboration activities and Faculty members' ratings is investigated. Two reasons lead to this investigation subsequent to the investigation of the main question: (1) If there were a substantial correlation, this would be a sign that quality in science is generally related to the number of researchers involved. Thus, not only citation impact – as a proxy of quality – would be affected by collaboration activities, but also other indicators of quality (e.g. expert assessments). (2) For the investigation of the relationship between collaboration and citation impact with quality being controlled for, a quality indicator is needed which is itself independent of collaboration activities. If it were dependent, the interpretation of the correlation coefficients between collaboration activities and citation impact would be difficult. Controlling for the members' ratings variable would also lead to controlling for collaboration activities in the regression model.

The results of the regression models show that the correlation between collaboration activities and Faculty members' ratings is weak. Thus, this indicator of quality is scarcely affected by the number of authors, the number of affiliations, and the number of countries given on a paper. From this one can infer that indicators of quality are not generally affected by collaborations or the quality of a paper does not depend on the number of scientists involved. Further, the Faculty members' ratings can be used as a relatively "unaffected" measure of quality (with regard to collaborations) in the main analysis of this study.

The main analysis focuses on the relationship between the number of authors, number of affiliations, and number of countries on a paper and its normalized citation score. The relationship is tested twice, with and without control for quality (measured by members' ratings). If the quality of a paper profited from a larger group of contributors, we could expect different results: With control for quality, the correlation between collaboration and citation impact should diminish as against the results without the control. As the results of the regression models show, the correlations between number of authors, number of affiliations, and number of countries on a paper and its normalized citation score are similar – independent of whether the Faculty members' ratings are included in the models or not. Thus, the increasing citation impact with collaboration activities seems to be rather citation specific and less quality related. More authors seem to mean more self-citations and a greater dissemination of papers to colleagues (Ronda-Pupo & Katz, 2015).

The possible lower relevance of quality for the citation impact advantage of papers written in collaboration is not only an interesting result in itself, but also has implications for the normalization of citation counts. Advanced bibliometrics does not use bare citation counts, but normalizes the citation counts of a paper with respect to its publication year and subject category (Bornmann, 2014b). This is the general practice in advanced bibliometrics because one can expect more citations for papers published in some subject categories (e.g. biomedicine) than in others (e.g. mathematics). Furthermore, one can expect fewer citations for papers published recently than for papers published long ago. The normalization based on both factors is only reasonable because both factors are not related to quality. If they were related to quality, the normalization would change the quality level of a paper from bare citation counts to normalized impact scores: the quality level would increase or decrease with the normalization of citation counts. Since this study shows that the number of authors, number of affiliations, and number of countries have an effect on citations which is scarcely quality related, these factors reflecting collaboration activities should also be considered in the normalization process of citation counts. It is primarily the citation practice which changes with the number of authors, affiliations, and countries but scarcely the quality of the papers.

What are the limitations of this study? (1) The first limitation concerns the main result of this study: the citation advantage of papers written in collaboration is scarcely related to their quality. Although this result suggests an effect of citation-specific factors (e.g. increased self-citations) on the citations of papers written in collaboration, it is not completely clear whether these factors are really essential here. Since the author of this paper does not have the necessary data at hand (e.g. the number of self-citations for the papers in the MPG in-house database or the size of authors' dissemination networks), future studies should investigate which citation-specific factors are significantly related to the citation advantage of papers written in collaboration. (2) Another limitation is the focus of this study on the biomedical area. Thus, the results are valid for this area and it is not clear whether they can be generalized. Therefore, it is desirable that the study is repeated in other areas (in a similar form). (3) This study uses Faculty members' ratings as an indicator for quality. However, these ratings are only proxies for quality, which might be biased (see above). The results of studies on the reliability of reviewers' recommendations in journal peer review processes show that reviewers frequently come to different conclusions on one and the same manuscript (Bornmann, Mutz, & Daniel, 2011). In order to increase the reliability of the quality assessments in this study, only papers with more than one members' ratings are included and the median is calculated over the different ratings per paper.

(3) The third limitation concerns the use of the F1000Prime post-publication peer review system as data source for this study. Several other studies have already investigated data from the system. Bornmann and Leydesdorff (2013), Li and Thelwall (2012), Mohammadi and Thelwall (2013), and Waltman and Costas (2014) calculated coefficients for the correlation between F1000Prime scores and citation impact. The meta-analysis of Bornmann (2015a) including the coefficients of these studies revealed a pooled r=.25 which can be interpreted as a medium effect size. The result of the meta-analysis with a relatively low pooled r might be a sign that F1000Prime scores are not able to reflect the concept of quality properly and are not representative of 'similar' journal peer review processes. Although this study is based on manuscripts with at least two scores by different Faculty members (to increase the reliability of the scores), the low inter-reviewer reliability revealed by Bornmann (2015a) also question the link between quality and F1000Prime scores.

Wardle (2010) concludes on the basis of empirical results: "If … the F1000 process is unable to identify those publications that subsequently have the greatest impact while highlighting many that do not, it cannot be reliably used as a means of post-publication quality evaluation at the individual, departmental, or institutional levels" (p. 14). Further weaknesses of the F1000Prime scores are that they depend on the Impact Factor (Garfield, 2006) of the journals, in which the papers have been published (Li & Thelwall, 2012), and are geographically biased (Wardle, 2010). In other words, the validity of the results in this study is restricted by the limitations of the F1000Prime dataset: It is not clear whether the quality of papers is reliable measured by Faculty members' scores.

Despite these weaknesses of the F1000Prime data, it is a unique dataset – with reviewer scores for a great many publications – for which alternative datasets scarcely exist. In the interpretation of empirical results based on the F1000Prime dataset the weaknesses should always be considered.

# 6 Conclusions

In recent years, the relationship of collaboration among scientists and the citation impact of papers have been frequently investigated. Most of the studies show that the two variables are closely related: increased collaboration activity (measured in terms of co-authorships) leads to increased citation impact. However, it is not clear whether the increased citation impact is especially associated with the higher quality of papers, which profit from more than one scientist giving expert input, or other factors. Thus, the current study addresses this question by using two comprehensive datasets based on publications (in the biomedical area) including quality assessments by experts (F1000Prime dataset) and citation data for the publications (data from an in-house database based on WoS). The matched dataset enables the investigation of the relationship between collaboration activities (measured in terms of number of authors, number of affiliations, and number of countries) and citation impact with the publications' quality (measured by the assessments of Faculty members) being controlled for.

Robust regression models are calculated in this study to investigate the relationship between number of authors, number of affiliations, and number of countries, respectively, and citation impact – controlling for the papers' quality (measured by F1000Prime expert ratings). The results point out that the effect of collaboration activities on citation impact is largely independent of the papers' quality. The citation advantage is apparently scarcely quality-related; citation specific factors (e.g. self-citations) seem to be important here. Thus, the results question the use of collaboration activities as a research performance indicator which is mainly based on the assumption that collaborations increase the quality of research per se. The problem is that research quality is conventionally measured by citation impact which is not only triggered by quality but also by self-citations and the authors' network activities (and possible other factors). Despite the limitations of the study which are especially related to the datasets used, the results of this study are an interesting contribution to the discussion concerning the relationship between collaboration in science and citation impact.

## Acknowledgements

I would like to thank Ros Dignon and Iain Hrynaszkiewicz from F1000 for providing me with the F1000Prime data set. The bibliometric data used in this paper are from an in-house database developed and maintained by the Max Planck Digital Library (MPDL, Munich) and derived from the Science Citation Index Expanded (SCI-E), Social Sciences Citation Index (SSCI), Arts and Humanities Citation Index (AHCI) prepared by Thomson Reuters.

# References

Acock, A. C. (2014). *A gentle introduction to Stata*. College Station, TX, USA: Stata Press.

Adams, J. (2012). Collaborations: The rise of research networks. *Nature, 490*(7420), 335-336.

Adams, J. (2013). Collaborations: the fourth age of research. *Nature, 497*(7451), 557-560. doi: 10.1038/497557a.

Bence, V., & Oppenheim, C. (2004). The influence of peer review on the Research Assessment Exercise. *Journal of Information Science, 30*(4), 347-368.

Berry, W. D., & Feldman, S. (1985). *Multiple regression in practice*. Thousand Oaks, CA, USA: Sage Publications.

Bidault, F., & Hildebrand, T. (2014). The distribution of partnership returns: Evidence from co-authorships in economics journals. *Research Policy, 43*(6), 1002-1013. doi: 10.1016/j.respol.2014.01.008.

Bornmann, L. (2011). Scientific peer review. *Annual Review of Information Science and Technology, 45*, 199-245.

Bornmann, L. (2014a). Do altmetrics point to the broader impact of research? An overview of benefits and disadvantages of altmetrics. *Journal of Informetrics, 8*(4), 895-903. doi: 10.1016/j.joi.2014.09.005.

Bornmann, L. (2014b). How are excellent (highly cited) papers defined in bibliometrics? A quantitative analysis of the literature. *Research Evaluation, 23*(2), 166-173. doi: 10.1093/reseval/rvu002.

Bornmann, L. (2014c). Validity of altmetrics data for measuring societal impact: A study using data from Altmetric and F1000Prime. *Journal of Informetrics, 8*(4), 935-950.

Bornmann, L. (2015a). Inter-rater reliability and convergent validity of F1000Prime peer review. *Journal of the Association for Information Science and Technology, 66*(12), 2415-2426.

Bornmann, L. (2015b). Usefulness of altmetrics for measuring the broader impact of research: A case study using data from PLOS and F1000Prime. *Aslib Journal of Information Management, 67*(3), 305-319. doi: 10.1108/AJIM-09-2014-0115.

Bornmann, L., & Haunschild, R. (2015). Which people use which scientific papers? An evaluation of data from F1000 and Mendeley. *Journal of Informetrics, 9*(3), 477–487.

Bornmann, L., & Leydesdorff, L. (2013). The validation of (advanced) bibliometric indicators through peer assessments: A comparative study using data from InCites and F1000. *Journal of Informetrics, 7*(2), 286-291. doi: 10.1016/j.joi.2012.12.003.

Bornmann, L., & Marx, W. (2014). The wisdom of citing scientists. *Journal of the American Society of Information Science and Technology, 65*(6), 1288-1292.

Bornmann, L., Mutz, R., & Daniel, H.-D. (2011). A reliability-generalization study of journal peer reviews - a multilevel meta-analysis of inter-rater reliability and its determinants. *PLoS One, 5*(12), e14331.

Bornmann, L., Stefaner, M., de Moya Anegón, F., & Mutz, R. (2015). Excellence networks in science: A web-based application (www.excellence-networks.net) for the identification of institutions collaborating successfully. Retrieved August, 2015, from URL fehlt noch

Bornmann, L., Wagner, C., & Leydesdorff, L. (2015). BRICS countries and scientific excellence: A bibliometric analysis of most frequently cited papers. *Journal of the Association for Information Science and Technology, 66*(7), 1507-1513. doi: 10.1002/asi.23333.

Bornmann, L., & Williams, R. (2013). How to calculate the practical significance of citation impact differences? An empirical example from evaluative institutional bibliometrics using adjusted predictions and marginal effects. *Journal of Informetrics, 7*(2), 562-574. doi: 10.1016/j.joi.2013.02.005.

Bozeman, B., Fay, D., & Slade, C. (2013). Research collaboration in universities and academic entrepreneurship: the-state-of-the-art. *The Journal of Technology Transfer, 38*(1), 1-67. doi: 10.1007/s10961-012-9281-8.

Cimenler, O., Reeves, K. A., & Skvoretz, J. (2014). A regression analysis of researchers' social network metrics on their citation performance in a college of engineering. *Journal of Informetrics, 8*(3), 667-682. doi: http://dx.doi.org/10.1016/j.joi.2014.06.004.

Cohen, J. (1988). *Statistical power analysis for the behavioral sciences* (2nd ed.). Hillsdale, NJ, USA: Lawrence Erlbaum Associates, Publishers.

Colledge, L. (2014). *Snowball Metrics Recipe Book*. Amsterdam, the Netherlands: Snowball Metrics program partners.

Elsevier and Science Europe. (2013). *Comparative benchmarking of European and US Research collaboration and researcher mobility*. Amsterdam, the Netherlands: Elsevier.

Garfield, E. (2006). The history and meaning of the Journal Impact Factor. *Journal of the American Medical Association, 295*(1), 90-93.

Glänzel, W. (2008). *Seven myths in bibliometrics. About facts and fiction in quantitative science studies.* Paper presented at the Proceedings of WIS 2008, Berlin. Fourth International Conference on Webometrics, Informetrics and Scientometrics & Ninth COLLNET Meeting, Berlin, Germany.

Haunschild, R., & Bornmann, L. (2015). F1000Prime: an analysis of discipline-specific reader data from Mendeley [version 2; referees: 1 approved with reservations, 1 not approved] *F1000Research, 4*(41). doi: 10.12688/f1000research.6062.2.

Jayasinghe, U. W. (2003). *Peer review in the assessment and funding of research by the Australian Research Council.* University of Western Sydney, Greater Western Sydney, Australia. Retrieved from http://handle.uws.edu.au:8081/1959.7/572

Katz, J. S., & Martin, B. R. (1997). What is research collaboration? *Research Policy, 26*(1), 1-18.

Kraemer, H. C., Morgan, G. A., Leech, N. L., Gliner, J. A., Vaske, J. J., & Harmon, R. J. (2003). Measures of clinical significance. *Journal of the American Academy of Child and Adolescent Psychiatry, 42*(12), 1524-1529. doi: 10.1097/01.chi.0000091507.46853.d1.

Kreiman, G., & Maunsell, J. H. R. (2011). Nine criteria for a measure of scientific output. *Frontiers in Computational Neuroscience, 5*(48). doi: 10.3389/fncom.2011.00048.

Kumar, S. (2015). Co-authorship networks: a review of the literature. *Aslib Journal of Information Management, 67*(1), 55-73. doi: 10.1108/ajim-09-2014-0116.

Larivière, V., Gingras, Y., Sugimoto, C. R., & Tsou, A. (2015). Team size matters: Collaboration and scientific impact since 1900. *Journal of the Association for Information Science and Technology, 66*(7), 1323-1332. doi: 10.1002/asi.23266.

Li, X., & Thelwall, M. (2012). F1000, Mendeley and traditional bibliometric indicators. In E. Archambault, Y. Gingras & V. Lariviere (Eds.), *The 17th International Conference on Science and Technology Indicators* (pp. 541-551). Montreal, Canada: Repro-UQAM.

Martin, B. R., & Irvine, J. (1983). Assessing basic research - some partial indicators of scientific progress in radio astronomy. *Research Policy, 12*(2), 61-90.

Marx, W., & Bornmann, L. (2015). On the causes of subject-specific citation rates in Web of Science. *Scientometrics, 102*(2), 1823-1827.

Merton, R. K. (1973). *The sociology of science: theoretical and empirical investigations*. Chicago, IL, USA: University of Chicago Press.

Milojević, S. (2014). Principles of scientific research team formation and evolution. *Proceedings of the National Academy of Sciences, 111*(11), 3984-3989. doi: 10.1073/pnas.1309723111.

Mohammadi, E., & Thelwall, M. (2013). Assessing non-standard article impact using F1000 labels. *Scientometrics, 97*(2), 383-395. doi: 10.1007/s11192-013-0993-9.

Ronda-Pupo, G. A., & Katz, J. S. (2015). The power–law relationship between citation-based performance and collaboration in articles in management journals: A scale-independent approach. *Journal of the Association for Information Science and Technology*, n/a-n/a. doi: 10.1002/asi.23575.

Schneider, J. W., & Sørensen, M. P. (2015). Measuring research performance of individual countries: the risk of methodological nationalism. Retrieved October 13, 2015, from http://pure.au.dk/portal/files/90990388/1bcd2793_df8a_42bc_92b8_722449962a7e.pdf

Sheskin, D. (2007). *Handbook of parametric and nonparametric statistical procedures* (4th ed.). Boca Raton, FL, USA: Chapman & Hall/CRC.

Smith, M. J., Weinberger, C., Bruna, E. M., & Allesina, S. (2014). The Scientific Impact of Nations: Journal Placement and Citation Performance. *PLoS ONE, 9*(10), e109195. doi: 10.1371/journal.pone.0109195.

Sonnenwald, D. H. (2007). Scientific collaboration. *Annual Review of Information Science and Technology, 41*(1), 643-681. doi: 10.1002/aris.2007.1440410121.

StataCorp. (2015). *Stata statistical software: release 14*. College Station, TX, USA: Stata Corporation.

Wagner, C. S., Park, H. W., & Leydesdorff, L. (2015). The Continuing Growth of Global Cooperation Networks in Research: A Conundrum for National Governments. *PLoS ONE, 10*(7), e0131816. doi: 10.1371/journal.pone.0131816.

Waltman, L., & Costas, R. (2014). F1000 recommendations as a potential new data source for research evaluation: a comparison with citations. *Journal of the Association for Information Science and Technology, 65*(3), 433-445.

Waltman, L., Tijssen, R. J. W., & van Eck, N. J. (2011). Globalisation of science in kilometres. *Journal of Informetrics, 5*(4), 574-582. doi: 10.1016/j.joi.2011.05.003.

Waltman, L., van Eck, N., van Leeuwen, T., Visser, M., & van Raan, A. (2011a). Towards a new crown indicator: an empirical analysis. *Scientometrics, 87*(3), 467-481. doi: 10.1007/s11192-011-0354-5.

Waltman, L., van Eck, N. J., van Leeuwen, T. N., Visser, M. S., & van Raan, A. F. J. (2011b). Towards a new crown indicator: some theoretical considerations. *Journal of Informetrics, 5*(1), 37-47. doi: 10.1016/j.joi.2010.08.001.

Wardle, D. A. (2010). Do 'Faculty of 1000' (F1000) ratings of ecological publications serve as reasonable predictors of their future impact? *Ideas in Ecology and Evolution, 3*, 11-15.

Weller, A. C. (2002). *Editorial peer review: its strengths and weaknesses*. Medford, NJ, USA: Information Today, Inc.

Wets, K., Weedon, D., & Velterop, J. (2003). Post-publication filtering and evaluation: Faculty of 1000. *Learned Publishing, 16*(4), 249-258.

Williams, R. (2012). Using the margins command to estimate and interpret adjusted predictions and marginal effects. *The Stata Journal, 12*(2), 308-331.

Williams, R., & Bornmann, L. (2014). The substantive and practical significance of citation impact differences between institutions: guidelines for the analysis of percentiles using effect sizes and confidence intervals. In Y. Ding, R. Rousseau & D. Wolfram (Eds.), *Measuring scholarly impact: methods and practice* (pp. 259-281). Heidelberg, Germany: Springer.

Wouters, P., & Costas, R. (2012). *Users, narcissism and control – tracking the impact of scholarly publications in the 21st century*. Utrecht, The Netherlands: SURFfoundation.
Wouters, P., Thelwall, M., Kousha, K., Waltman, L., de Rijcke, S., Rushforth, A., & Franssen, T. (2015). *The Metric Tide: Correlation analysis of REF2014 scores and metrics (Supplementary Report II to the Independent Review of the Role of Metrics in Research Assessment and Management)*. London, UK: Higher Education Funding Council for England (HEFCE).
Ziman, J. (2000). *Real science. What it is, and what it means*. Cambridge, UK: Cambridge University Press.